# Public awareness and attitudes towards search engine optimization

Dirk Lewandowski & Sebastian Schultheiß
Hamburg University of Applied Sciences
Faculty Design, Media and Information
Department of Information
Finkenau 35
22081 Hamburg
Germany
dirk.lewandowski@haw-hamburg.de, sebastian.schultheiss@haw-hamburg.de

**Abstract**

This research focuses on what users know about search engine optimization (SEO) and how well they can identify results that have potentially been influenced by SEO. We conducted an online survey with a sample representative of the German online population ($N$ = 2,012). We found that 43% of users assume a better ranking can be achieved without paying money to Google. This is in stark contrast to the possibility of influence through paid advertisements, which 79% of internet users are aware of. However, only 29.2% know how ads differ from organic results. The term "search engine optimization" is known to 8.9% of users but 14.5% can correctly name at least one SEO tactic. Success in labelling results that can be influenced through SEO varies by search engine result page (SERP) complexity and devices: participants achieved higher success rates on SERPs with simple structures than on the more complex SERPs. SEO results were identified better on the small screen than on the large screen. 59.2% assumed that SEO has a (very) strong impact on rankings. SEO is more often perceived as positive (75.2%) than negative (68.4%). The insights from this study have implications for search engine providers, regulators, and information literacy.

# Introduction

Search engines on the web shape the online experience of billions of users worldwide. Search is ubiquitous and taken for granted. It has become so commonplace, some argue we no longer even notice (Haider & Sundin, 2019) that search exists as a service. Search engines provide users with ranked search outputs from which they usually select only one site to visit or, in some cases, a handful. These search results influence user opinions and decisions. For instance, the order of search engine results alone can sway a user's choice of political candidate (Epstein & Robertson, 2015). Because search engines have become a primary means of finding information, being visible on search engine result pages (SERPs) is of pivotal importance to businesses, institutions, and other organizations. In the context of recent discussions of "information disorders" (Lazer et al., 2018) such as fake news, misinformation, and the "infodemic" (Xie et al., 2020), it has been argued that a plethora of misinformation exists on social media and the web. Worse yet, research has found that users more frequently share false information than correct information (Vosoughi et al., 2018). International organizations including the United Nations and the World Health Organization have announced measures to prevent the spread of misinformation (United Nations, 2020; World Health Organization, 2020). Although it remains unclear which measures are most effective, fact-checking approaches or technical measures alone do not sufficiently combat misinformation (Lewandowsky et al., 2017).

It is striking that the discussion on information disorders has primarily focused on social media, while information acquisition through search engines has not been considered in depth. There is an ongoing discussion on search engine bias, or how search engines may prefer specific results over others in an unfair manner (e.g., Gezici, Lipani, Saygin, & Yilmaz, 2021; Noble, 2018; Otterbacher, Bates, & Clough, 2017). This discussion is, however, more on a conceptual level and includes at best empirical studies using query sets assembled through convenience sampling. Large-scale empirical studies like the ones conducted on social media are lacking.

It is imperative to understand how search engines may favor some results and thereby influence what users get to see on the search engine result pages (SERPs; Goel et al., 2010; Höchstötter & Lewandowski, 2009; Lewandowski et al., 2021; Lewandowski & Sünkler, 2019), which results they (visually) pay attention to (Lewandowski & Kammerer, 2020; Strzelecki, 2020), and which results they select (e.g., Pan et al., 2007; Schultheiß et al., 2018). We must, however, also obtain a basic understanding of the degree to which users understand what is presented on the SERPs, which interests influence those results, and what users think about these influences when they are in fact aware of them.

In research and development of search systems in the field of information retrieval (Büttcher & Clarke, 2010; Levene, 2011; Manning et al., 2008), a fundamental assumption is that the sole aim of search systems is to produce relevant results (system-side relevance). Furthermore, it is assumed that users select the results most relevant to them from a list presented in response to their queries (user-side relevance) (Saracevic, 2016). These assumptions, however, neglect another fundamental layer of relevance, namely to society (Haider & Sundin, 2019). Commercial interests that affect the results of search engines (see also Schultheiß & Lewandowski, 2021b) also warrant consideration.

One type of commercial interest arises from the self-interests of search providers including the promotion of content from subsidiaries and their own vertical search engines such as specialized news and video search engines. In 2017, the European Commission fined Google 2.4 billion Euros (European Commission, 2017) for unfairly promoting its own shopping search results, the highest fine to that date. Even more importantly, they forced Google to take measures to make the way it presents results fair.

Keyword-based advertising represents another commercial interest at play. Contextual advertisements ("sponsored results") are shown in response to user queries, usually above the list of what are referred to as organic results (Jansen, 2011). Since these ads are generated in response to user queries, they should be considered a type of search result (Lewandowski et al., 2018). Research has shown that a large proportion of users is unable to distinguish between organic results and ads (Lewandowski et al., 2018), and that users who lack digital advertising knowledge select the first ad that is displayed more than twice as often as knowledgeable users

(Lewandowski, 2017b). Search engine companies generate the overwhelming majority of their revenue by selling ads. For instance, Alphabet Inc., the parent company of Google, makes 83 percent of its revenue through advertising (Alphabet Inc., 2020). Therefore, search engine companies may be tempted to blur the lines between advertisements and organic results to increase their revenues from clicks on ads (Lewandowski et al., 2018; Schultheiß & Lewandowski, 2021a).

A third type of commercial interest involves content providers and companies that sell products or services over the internet. They conduct search engine optimization (SEO) in an attempt to make their content more visible in search results (Enge et al., 2015; Moran & Hunt, 2015; Thurow, 2007). Search engine optimization is "the practice of optimizing web pages in a way that improves their ranking in the organic search results" (Kai Li et al., 2014). The market for SEO has reached 80 billion dollars in the U.S. alone (McCue, 2018). We can reliably assume that SEO has a significant influence on search results (see Lewandowski et al., 2021) and that search engine users select from results that are not only displayed because of their relevance, but also because they have been optimized to rank higher in the search engine.

Below, we examine via an online survey using representative quota sampling perceptions of search engine optimization (SEO) among users and the extent to which they can identify content that may be influenced by SEO on the search results pages of Google, the most popular search engine on the web. Our goal is to contribute to a better understanding of people's knowledge on how search results are produced in response to a complex interaction between the interests of search engine providers and content providers, the latter mediated through search engine optimization. To fully grasp how commercial interests and manipulative practices (the dissemination of misinformation, disinformation, and fake news) are influencing knowledge and beliefs — and to design effective regulatory interventions — it is crucial to first understand how aware users are of the effect SEO has on search results.

## Literature review

### The role of search engines

Search engines shape what billions of people see as they pursue their information needs by formulating search queries. Search is an integral part of being an informed citizen in the modern world. Therefore, as Vint Cerf recently put it, "it seems inescapable that the presentation of search results not only must be prioritized by some measure of quality but also that the ranking criteria must be clear and well understood." (Cerf, 2019). However, several studies auditing search engine algorithms have shown that this is not the case. On the system side, it has been demonstrated that search engines such as Google in many cases show biased results (Ciampaglia et al., 2018; Gao & Shah, 2020), for instance in terms of race or gender (Kulshrestha et al., 2019; Noble, 2018; Otterbacher et al., 2017), and that they promote hate speech (Noble, 2012, 2018) and conspiracy theories (Ballatore, 2015). On the user side, research has shown that users only have a limited understanding of search engines and their ranking criteria (European Commission, 2016; Lewandowski et al., 2018; Purcell et al., 2012).

### Trust in search engines

Users place great trust in search engines. This is reflected by the 91% of US users who said they find what they are looking for always or most of the time, and the 66% who believe search engines are a fair and unbiased source of information (Purcell et al., 2012). Furthermore, 78% of European internet and online platform users said they trust that their search engine results are the most relevant results (European Commission, 2016). Globally, users trust search engines more than any other source (including traditional news outlets) when it comes to news (Edelman, 2019), and users trust news found via search significantly more than news found on social media (Newman et al., 2020). Trust is further expressed through users relying on the top results presented (Bar-Ilan et al., 2009; Pan et al., 2007; Schultheiß et al., 2018; Westerwick, 2013). User behavior can often be characterized as "satisficing" (Simon, 1955) — people stop searching as soon as they find information that is 'good enough' (Bawden & Robinson, 2009). Together with the distribution of domains shown in search engine results being highly skewed (Goel et al., 2010), this leads to users selecting only from a limited set of sources. For instance, a study analyzing 2.5 billion queries from the Yahoo search engine found that 10,000 websites account for approx. 80 percent of user clicks (Goel et al., 2010). This shows the huge influence one can gain by appearing at the top of search engine results, especially in the topmost position. In aggregate, this means the ordering of results may not only be used to influence which products and services users buy but also to manipulate voter turnout and vote choice (Aral & Eckles, 2019), thereby influencing political elections (Epstein & Robertson, 2015). It is important to stress that even minor manipulations (e.g., switching two result positions) may have a considerable effect on how users perceive information on SERPs and which results they select (Epstein & Robertson, 2015).

## Search engine optimization

Search engine optimization is part of search engine marketing (SEM). In addition to SEO, SEM also includes paid search marketing (PSM), which refers to keyword-related advertisements (Kai Li et al., 2014) known as sponsored links. To be shown on the SERPs, advertisers pay the search engine for each click the ad attracts. Because these ads represent the primary source of revenue for search engines, many queries output at least two result lists on the search engine result page: A list of advertisements (paid for by external parties) and a list of organic results (that are not paid, but may have been optimized using SEO techniques).

Search engine traffic is important for many businesses, including media outlets (for traffic statistics, see similarweb.com, 2020). For content providers, gaining traffic through unpaid organic results is an attractive proposition. Although they must invest in optimizing their websites, there is no direct payment to the search engine involved. In 2020, SEO revenue is expected to reach $80 billion in the U.S. alone (McCue, 2018). Research indicates that many of the results shown on the first page of Google results for a query have been optimized through SEO techniques (Lewandowski et al., 2021).

Search engine optimization not only influences how websites are designed and how content is presented. Its influence begins as early as the content production stage. For instance, as SEO is essential for news outlets, it is already part of the curriculum of journalism programs (Lopezosa et al., 2020). SEO increasingly influences how journalists write, and SEO policies are applied in newsrooms (Giomelakis et al., 2019). In studies, interviewed journalists have expressed reservations about SEO in the context of journalistic work (Dick, 2011; Giomelakis et al., 2019; Prawira & Rizkiansyah, 2018). Giomelakis et al. (2019) found that, while SEO was considered indispensable as it ensures visibility of the content, it also significantly influences topic setting. This, in turn, can reduce the quality of the journalism being produced.

SEO can have both positive and negative effects from the perspective of the user and the search engine. A positive correlation exists between SEO tactics and website usability (Visser & Weideman, 2011) and accessibility (Moreno & Martinez, 2013). Further positive effects may come from website owners preparing their content in a way that allows users to find it easily. Search engine ranking algorithms factor in user satisfaction metrics (Diaz, 2016). Consequently, one objective of SEO is to enhance user satisfaction, with a corresponding positive influence on the user experience. The positive effects of SEO are also stressed by SEO professionals themselves, who argue they are helping make content discoverable and bring good content into the top results of search engines (Schultheiß & Lewandowski, 2020). While SEO undoubtedly has these positive effects, it is unclear whether SEO also improves the relevance of search results. It is assumed that vast amounts of low-quality content are produced to satisfy search engine algorithms, being created based on keyword research. As it is easier to produce low-quality content, and many more people produce it, there is much more low-quality content on the web. This so-called data bias may lead to search engines reproducing what there is on the web and therefore leaning towards low-quality content (Baeza-Yates, 2018). In this environment, actors interested in producing and distributing misinformation and fake news can thrive by exploiting data voids.

This strategy involves achieving the top-ranking result for less popular queries in an attempt to exploit the trust that users have in search engines (Golebiewski & Boyd, 2019). Conspiracy theorists, for example, have been known to bolster the legitimacy of their claims by pointing to high-ranking results in search engines (Ballatore, 2015). The prevailing high level of trust in search engines with users who behave accordingly together with the fact that a single search engine currently dominates the market gives us a situation in which a top search result can lead to huge traffic gains for a content provider, but also cause users to select the linked content (European Commission, 2016; Purcell et al., 2012; Westerwick, 2013).

Given the enormous amount of money invested in search engine optimization and its huge potential to influence knowledge acquisition through search engines, it is surprising that little research has been conducted on the effects of SEO. The literature primarily focuses on practical strategies for optimizing websites (e.g., Enge et al., 2015; Moran & Hunt, 2015; Thurow, 2007) and attempts to identify ranking factors used by commercial search engines and exploit this knowledge to boost rankings (Barbar & Ismail, 2019; Drivas et al., 2020; Evans, 2007; Giomelakis & Veglis, 2016; Hoyos et al., 2019; Su et al., 2014; Umenhofer, 2019; Ziakis et al., 2019). Also, some research has investigated the role of SEO professionals (Schultheiß & Lewandowski, 2020; Ziewitz, 2019; Zuze & Weideman, 2013) and SEO in professional contexts such as news production (Giomelakis et al., 2019). However, there is a significant research gap surrounding the impact of search engine optimization on users.

## Research questions

The literature review showed that search engines play a tremendously important role in knowledge acquisition and that users overwhelmingly trust search engines to provide relevant information. Results presented on SERPs are, however, influenced by internal and external factors: search engine providers are pursuing their own interests with their result pages, and external actors are trying to gain visibility via search engine optimization. A link appearing among highly ranked results confers legitimacy. This leads to a situation in which users may place greater trust in the credibility of results included on a SERP than is in fact warranted. To actors aware of this situation, it represents an opportunity to spread misinformation and fake news via SEO techniques.

One major problem is that search engine optimization usually cannot be detected by merely examining web pages, except for very shallow SEO attempts such as optimizing for a fixed keyword density. Optimized results are not labeled, unlike advertisements, which carry a small ad label or the like. We therefore hypothesized that users are even less familiar with SEO than they are with paid search marketing. With our research, we aim to quantify user knowledge of search engine optimization, gain insight into user attitudes towards SEO, and understand the potential consequences for knowledge acquisition over the internet. The following research questions were derived from the literature referred to above and the results of interviews we conducted with SEO experts (Schultheiß & Lewandowski, 2020):

- RQ1: What knowledge do internet users have of paid search marketing (PSM)?
- RQ2: What knowledge do internet users have of search engine optimization (SEO)?
- RQ3: Are internet users able to distinguish between organic results and ads on Google's search engine results pages?
- RQ4: How strong do internet users believe SEO can impact the search results?
- RQ5: What opinions do internet users have about the influence that SEO can have on search results?

The primary aim of our study lies in tackling the research questions on SEO (RQ2, RQ4, RQ5), which have not been addressed in prior research. To provide our research on SEO in the proper context, we use two research questions (RQ1 and RQ3) that have already been tackled before but where there is no current data, which we need for comparison reasons. We decided to frame these as separate research questions as they have only been addressed in a single study, and our results will strengthen its evidential value.

## Methods

### Data collection

We conducted a representative online survey with *N* = 2,012 German internet users. The sample is representative of the German online population aged 16 through 69, according to the criteria applied by "Arbeitsgemeinschaft Onlineforschung" (Working Group Online Research; AGOF)[1]. The survey was carried out with the market research company Fittkau & Maaß Consulting between January and April 2020. The market research firm suggested a sample size of about *N* = 1,000 subjects. However, since the subjects completed our study either on a large or on a small screen (see section Marking tasks), we formed two subsamples of *N* = 999 subjects (large screen) and *N* = 1,013 subjects (small screen), which meet the same requirements regarding representativeness described above. Subjects who opted out of the questionnaire or solely clicked through were replaced by the market research firm. Each subject received EUR 0.75 for complete participation. See Table 1 for demographic data on the sample.

---

[1] https://www.agof.de/en/

*Table 1: Demographic information*

|  | Large screen | | Small screen | |
|---|---|---|---|---|
| **Baseline characteristic** | *n* | *%* | *n* | *%* |
| Sample size | 999 | 100 | 1013 | 100 |
| **Age** | | | | |
|   16 to 24 | 134 | 13.4 | 145 | 14.3 |
|   25 to 34 | 193 | 19.3 | 203 | 20.0 |
|   35 to 44 | 209 | 20.9 | 208 | 20.5 |
|   45 to 54 | 245 | 24.5 | 243 | 24.0 |
|   55 to 69 | 218 | 21.8 | 214 | 21.1 |
| **Gender** | | | | |
|   Female | 488 | 48.8 | 492 | 48.6 |
|   Male | 511 | 51.2 | 521 | 51.4 |
| **Education** | | | | |
|   Certificate of Secondary Education without completed apprenticeship | 24 | 2.4 | 22 | 2.2 |
|   Certificate of Secondary Education with completed apprenticeship | 102 | 10.2 | 109 | 10.8 |
|   General Certificate of Secondary Education | 298 | 29.8 | 301 | 29.7 |
|   A-levels (university entrance qualification) | 286 | 28.6 | 295 | 29.1 |
|   University degree | 253 | 25.3 | 244 | 24.1 |
|   None | 0 | 0.0 | 1 | 0.1 |
|   (Still) without school-leaving certificate (e.g., student) | 8 | 0.8 | 16 | 1.6 |
|   Other | 28 | 2.8 | 25 | 2.5 |
| **Preferred search engine** | | | | |
|   Google | 883 | 89.2 | 907 | 90.4 |
|   Ecosia | 51 | 5.2 | 61 | 6.1 |
|   DuckDuckGo | 24 | 2.4 | 15 | 1.5 |
|   Bing | 17 | 1.7 | 16 | 1.6 |
|   Web.de | 8 | 0.8 | 1 | 0.1 |
|   Yahoo! | 3 | 0.3 | 2 | 0.2 |
|   Another | 4 | 0.4 | 0 | 0.0 |
|   I don't know/not specified | 0 | 0.0 | 1 | 0.1 |

## Study design

The survey had five sections. These are, as shown in Figure 1, (I) Public awareness of paid search marketing (PSM), (II) Public awareness of search engine optimization (SEO), (III) Public attitudes towards SEO, (IV) Identification of the areas of influence of SEO and PSM, and (V) Demographic data and search engine preferences.

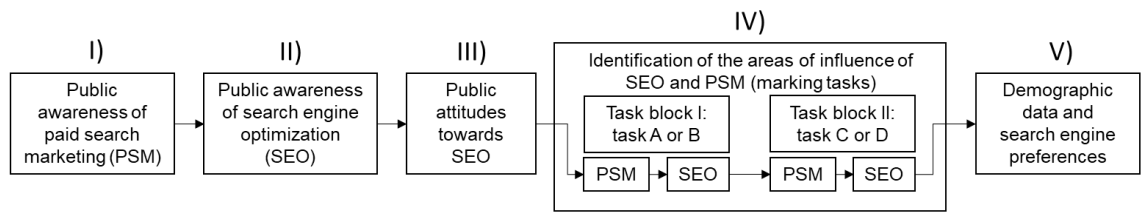

*Figure 1: Flow chart of the study procedure*

In the first section (I), we asked questions about users' understanding of paid search marketing (PSM), which we adopted from the study by Lewandowski et al. (2018). In section II, we focused on the knowledge of internet users about search engine optimization. For this purpose, we asked both open and closed questions that gradually approached the topic of SEO. This was done to determine whether users know that it is possible to influence search results apart from PSM. In addition to user awareness of SEO, we were also interested in opinions regarding the influence SEO can have on results, which we asked about in section III. Before proceeding with section III, we offered brief explanations[2] of SEO and PSM to ensure that all respondents could understand the survey tasks that followed. In section IV, we examined how accurately the respondents can assign search results to the potential influences (SEO, PSM). For this purpose, we used screenshot-based marking tasks, which we describe in the next section. At the end of the survey, we collected respondents' socio-demographic data (age, gender, highest educational level, and current occupation) and their search engine preferences.

The complete online survey includes sections that go beyond the focus of this paper and are reported elsewhere (Schultheiß & Lewandowski, 2021b). The survey was conducted in German. An English translation of the survey questions relevant to the research reported in this paper is shown in Table 2.

*Table 2: Questionnaire*

| No. | Question | Response options and correct answers/coding specifications |
|---|---|---|
| **I) Public awareness of paid search marketing (PSM)** | | |
| 1.1* | What do you think: Where does Google generate most of its revenue from? | free input<br>- correct: "ads" or terms having the same meaning<br>- partly correct: correct term and at least one incorrect term<br>- incorrect: clearly incorrect terms (e.g., data sale, donations) |
| 1.2* | Do website operators or companies have the opportunity to pay for their results to appear high up on Google's search results page? | yes (correct), no, I don't know |
| 1.3* | [If "Yes" on question 1.2]: Do such paid search results differ from the other search results? | yes (correct), no, I don't know |
| 1.4* | [If "Yes" on question 1.3]: And how do the paid search results on Google differ from the other results that have not been paid for? | free input<br>- correct: "ad label" or terms having the same meaning<br>- partly correct: correct term and at least one incorrect term<br>- incorrect: clearly incorrect terms (e.g., different font) |
| **II) Public awareness of search engine optimization (SEO)** | | |
| 2.1 | Do website operators or companies have the ability or influence to appear higher in the Google results list for certain queries without paying any money to Google? | yes (correct), no, I don't know |
| 2.2 | [If "Yes" on question 2.1]: Do you know what term is used to describe these measures to improve the ranking in the Google search results list (without payment to Google)? | free input<br>- correct: "search engine optimization" or abbreviation/German form<br>- partly correct: correct term and at least one incorrect term<br>- incorrect: clearly incorrect terms (e.g., ads) |
| 2.3 | [If "Yes" on question 2.1]: And by what means can a website be designed or programmed so that it is ranked higher in the Google search results lists? | free input<br>- correct: "keywords" or other correct SEO techniques<br>- partly correct: correct term and at least one incorrect term<br>- incorrect: clearly incorrect SEO techniques (e.g., payment) |
| | *Information part "SEO/PSM":*<br>*Website operators have several ways to ensure that their web pages appear at the top of the Google result page for a specific query, namely I) Payment: They pay money to Google**, or II) Search engine optimization: They design their websites accordingly, e.g., by using certain keywords, quick page speed, and appropriate image titles and descriptions. Next, we will show you two different Google result pages and would like to ask you which results can be influenced by payment to Google and/or search engine optimization.* | |
| **III) Public attitudes towards search engine optimization** | | |
| 3.1 | Now please think again about search engine optimization. In your opinion, how strong is the influence of search engine optimization on the ranking of the search results in Google? | very strong influence, major influence, medium influence, little influence, no influence, I don't know |
| 3.2 | How big are the positive effects of search engine optimization on the Google search results from your perspective? | very large, large, medium, low, non-existent, I don't know |
| 3.3 | How big are the negative effects of search engine optimization on the Google search results from your perspective? | |

---

[2] See research data, available at https://doi.org/10.17605/OSF.IO/JYV9R, for details.

| No. | Question | Response options and correct answers/coding specifications |
|---|---|---|
| 3.4 | [Respondents who assumed large or very large positive effects***]: Which positive effects does search engine optimization have in your opinion? | free input, I don't know |
| 3.5 | [Respondents who assumed large or very large negative effects***]: Which negative effects does search engine optimization have in your opinion? | |
| **IV) Identification of the areas of influence of SEO and PSM** | | |
| 4.1* | You will now see a Google result page. Are there any search results on this page that can be influenced by the website operator paying Google? | Marking the requested results or skipping the task by specifying that the requested result type is not available on SERP. |
| 4.2 | One more question about this search result page. Are there any search results on this page that can be influenced by search engine optimization? | |
| 4.3* | You will now see another Google result page. Are there any search results on this page that can be influenced by the website operator paying Google? | |
| 4.4 | One more question about this search result page. Are there any search results on this page that can be influenced by search engine optimization? | |

*\* Tasks based on (Lewandowski et al., 2018)*
*\*\* This is a simplified explanation. The fact that a payment to Google is only made after an ad is selected was left unmentioned for the sake of comprehensibility.*
*\*\*\* We assumed these subjects were most likely to respond to the open follow-up questions.*

## Marking tasks

For the marking tasks, we created two blocks with a total of four tasks to address a variety of SERP elements. Tasks A and B were assigned to block I (simple SERPs), tasks C and D to block II (complex SERPs). The structure of the two SERPs per block is identical in terms of their elements as shown in Table 3.

*Table 3: Marking tasks: queries and SERP elements*

| Block | Task | Query (translated) | SERP elements |
|---|---|---|---|
| I (simple) | A | tax return help | Organic results, |
| | B | legal advice | Text ads (top and bottom of SERP) |
| II (complex) | C | apple iphone | Organic results, |
| | D | samsung galaxy | Text ads (top of SERP), Shopping ads, News, Knowledge Graph |

Since we tested all tasks on large and small screens, we created eight SERP screenshots. Each participant received two screenshots, one randomly assigned from block I and one from block II (e.g., SERP screenshots for tasks A and C). Each SERP was shown two times. First, all PSM results were to be marked (i.e., text ads, shopping ads), and second, all SEO results[3] (organic results, news). When clicking on a result, it was highlighted in blue and thus marked as PSM or SEO result according to the task. By clicking on the same result again, the marking could be removed. Figure 2 shows a section of the large screen SERP of task C ("apple iphone"). The PSM results are indicated by a dashed line, the SEO results by a solid line.

---

[3] Please note that "SEO results" in the following refers to *all* organic results, regardless of whether the corresponding websites actually perform SEO.

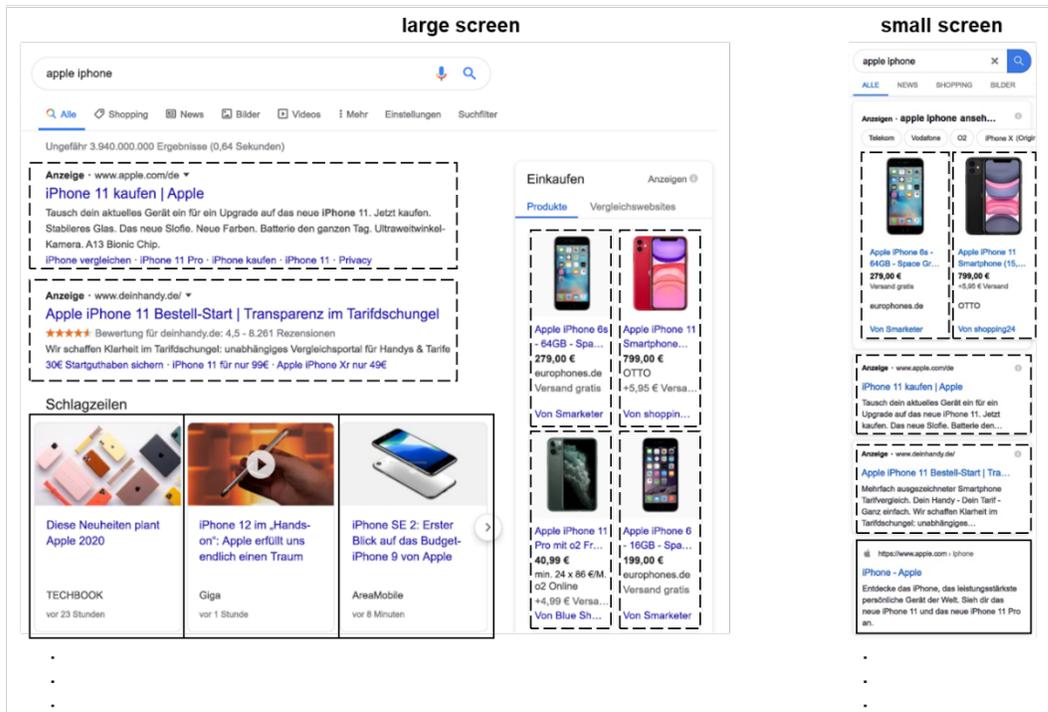

*Figure 2: Section of SERPs (large screen: left, small screen: right) for task C (PSM results: dashed line, SEO results: solid line)*

We used the desktop version of the Chrome browser and extensions for window resizing[4], user-agent switching[5], and screen capturing[6] to create the screenshots. The small-screen SERP screenshots were matched with the large-screen screenshots regarding results shown and their respective positions. Otherwise, different marking behavior might not be due to screen size differences (large vs. small) but due to varying results at different positions. An overview of all screenshots can be found in Appendix 1 (large screen) and Appendix 2 (small screen).

## Success rate for marking tasks

Success rates for each participant were calculated from the markings, SERP, screen size, and task (PSM/SEO). This rate considers correctly (true positive) and incorrectly (false positive) marked results according to the formula $\frac{n\ true\ -\ n\ false}{n\ to\ be\ marked}$ .

Two examples follow, the first for achieving a positive, the second for achieving a negative success rate. For SERP B, large screen, SEO results, ten organic results are to be marked, of which an example subject marks four (4 true). In addition, they incorrectly mark two ads (2 false). This results in a success rate of 0.2. Negative success rates are possible if subjects make more false than correct markings, e.g., for SERP A, small screen, PSM results, where four text ads are to be marked. If all of them are identified (4 true), but additionally, six organic results are marked (6 false), the subject would have a success rate of -0.5.

---

[4] https://chrome.google.com/webstore/detail/window-resizer/kkelicaakdanhinjdeammmilcgefonfh
[5] https://chrome.google.com/webstore/detail/user-agent-switcher-for-c/djflhoibgkdhkhhcedjiklpkjnoahfmg
[6] https://chrome.google.com/webstore/detail/full-page-screen-capture/fdpohaocaechififmbbbbbknoalclacl

### Grouping of participants

In the results section, we categorize subjects by the factors "age," "educational background," and "SEO affinity." These distinctions are made because interviews with experts indicated differences in user perspectives on SEO may exist that correlate with these factors (Schultheiß & Lewandowski, 2020). We used the respondents' demographic data to split them into the following age groups: "under 30," "30 to under 50," and "50 years and older." Level of education was divided into two groups since the interviewed experts (Schultheiß & Lewandowski, 2020) explicitly assumed there would be differences between people with a high school diploma or university degree and those who do not have such qualifications. Table 4 shows how the different types of professional activities, training, and studies correspond to affinity for SEO topics (low, average, high). To group the topics, we examined the module handbooks of the studies for intersections with the SEO topic. In the case of training and professional activity, e.g., pedagogy, we examined corresponding studies, e.g., educational science.

*Table 4: Affinity to SEO topics (grouping for professionals and students)*

| Topics | Affinity to SEO |
|---|---|
| **Professional activity** | |
| Purchasing, procurement, logistics | low |
| Finance, controlling | low |
| Production, manufacturing | low |
| Law | low |
| Marketing, sales, distribution | average |
| IT | average |
| Digitalization, internet | high |
| E-commerce, online trading | high |
| Online marketing, social media | high |
| **Training/studies** | |
| Business studies or economics | low |
| Engineering, electrical engineering | low |
| Law | low |
| Pedagogy | low |
| Social sciences | low |
| Informatics, business informatics | average |
| Digitalization, internet | high |
| E-commerce, online trading | high |
| Online marketing, social media | high |

### Statistical methods

Chi-square tests of independence, including the calculation of effect sizes, were performed to examine the relationship between (1) the knowledge on SEO/PSM and (2) the user groups regarding age, educational background, and SEO affinity.

## Results

The following presentation of results is split into three parts: First, we report public awareness of results influenced by commercial interests through paid search marketing and search engine optimization. Then, we report how well users can identify areas of SERPs that can be influenced either through SEO or PSM. Finally, we report results on public attitudes towards search engine optimization.

## Public awareness of commercially influenced results

Below we present results on how aware users are that results can be influenced by external actors, in other words, PSM and SEO.

**Public awareness of paid search marketing (PSM)**

In response to the question regarding Google's primary source of revenue (Q1.1), an at least partially correct answer was given by 68% of internet users. As a result, 32% of users do not know how Google generates its revenue — they either claim not to know Google's main revenue source (26%) or answer incorrectly (6%; n=112). A closer look at the incorrect responses reveals that most often, the sale of user data was mentioned (2.0%; n=41), followed by other data-related statements (1.7%; n=34).

The other questions (Q1.2-Q1.4) build on one another. The results to the question on whether website operators have the option of paid placements (Q1.2) show that 79% of internet users are aware of this possibility. In a follow-up question, these users were asked whether organic results can be distinguished from paid results, which was affirmed by 42% of all subjects (Q1.3). In turn, these users were asked how paid results differ from organic results (Q1.4). A total of 29% of all internet users provided a correct answer, such as "ad label."

*Table 5: Public awareness of PSM*

|  | Internet users | | Age | | | | | | Educational level | | | | SEO affinity | | | | | |
|---|---|---|---|---|---|---|---|---|---|---|---|---|---|---|---|---|---|---|
|  | | | under 30 years of age | | 30 to under 50 years of age | | 50 years and older | | A-levels or University degree | | lower | | high | | average | | low | |
|  | n | % | n | % | n | % | n | % | n | % | n | % | n | % | n | % | n | % |
| **Q1.1** | | | | | | | | | | | | | | | | | | |
| **(Partly) correct** | 1375 | 68 | 358 | 70 | 571 | 68 | 446 | 68 | 821 | 76 | 554 | 60 | 145 | 77 | 289 | 77 | 915 | 69 |
| **Incorrect/I don't know** | 637 | 32 | 152 | 30 | 273 | 32 | 210 | 32 | 263 | 24 | 372 | 40 | 43 | 23 | 85 | 23 | 413 | 31 |
| **Q1.2** | | | | | | | | | | | | | | | | | | |
| **Yes** | 1591 | 79 | 426 | 84 | 669 | 79 | 495 | 75 | 946 | 87 | 645 | 70 | 165 | 88 | 327 | 87 | 1062 | 80 |
| **No/I don't know** | 419 | 21 | 83 | 16 | 178 | 21 | 158 | 24 | 138 | 13 | 281 | 30 | 23 | 12 | 48 | 13 | 263 | 20 |
| **Q1.3** | | | | | | | | | | | | | | | | | | |
| **Yes** | 852 | 42 | 253 | 50 | 387 | 46 | 211 | 32 | 551 | 51 | 301 | 33 | 106 | 56 | 193 | 52 | 583 | 44 |
| **No/I don't know** | 739 | 37 | 173 | 34 | 283 | 34 | 284 | 43 | 395 | 36 | 343 | 37 | 59 | 31 | 134 | 36 | 479 | 36 |
| **Question not received** | 419 | 21 | 83 | 16 | 178 | 21 | 158 | 24 | 138 | 13 | 281 | 30 | 23 | 12 | 48 | 13 | 263 | 20 |
| **Q1.4** | | | | | | | | | | | | | | | | | | |
| **(Partly) correct** | 587 | 29 | 175 | 34 | 274 | 32 | 138 | 21 | 414 | 38 | 172 | 19 | 78 | 41 | 145 | 39 | 408 | 31 |
| **Incorrect/I don't know** | 269 | 13 | 77 | 15 | 115 | 14 | 76 | 12 | 137 | 13 | 132 | 14 | 29 | 15 | 49 | 13 | 177 | 13 |
| **Question not received** | 1154 | 57 | 258 | 51 | 455 | 54 | 442 | 67 | 533 | 49 | 622 | 67 | 81 | 43 | 180 | 48 | 743 | 56 |

There are significant differences in age, educational background, and affinity to most of the PSM topics, as the analyses (Table 6) for the data listed in Table 5 show. The lower the age and the higher the level of education and SEO affinity, the more frequently respondents provided a correct answer to the PSM questions. As Table 6 shows, correlations with the variables "PSM knowledge" (ratio of correct to incorrect answers for each question) and "educational background" are significant for all questions on PSM, with small effect sizes. Related to the other two factors (age, SEO affinity), we found non-significant results for three questions: Google's revenue source is equally familiar to all age groups (Q1.1). The fact that paid results differ from non-paid results is equally familiar to the different SEO affinity groups (Q1.3). Google's marking of ads is equally known to the different age and SEO affinity groups (Q1.4).

*Table 6: Results of the chi-square tests for the questions on PSM*

| Question No. | Age (3 groups) | Educational background (2 groups) | SEO affinity (3 groups) |
|---|---|---|---|
| 1.1 | ($x^2$(2, $N$ = 2010) = 1.030, $p$ = .597) | ($x^2$(1, $N$ = 2010) = 58.498, $p$ < .001), small effect size ($\varphi$ = 171, $p$ < .001) | ($x^2$(2, $N$ = 1890) = 13.392, $p$ = .001), small effect size ($\varphi$ = 084, $p$ = .001) |
| 1.2 | ($x^2$(2, $N$ = 2009) = 10.809, $p$ = .004), small effect size ($\varphi$ = .073, $p$ = .004) | ($x^2$(1, $N$ = 2010) = 93.911, $p$ < .001), small effect size ($\varphi$ = .216, $p$ < .001) | ($x^2$(2, $N$ = 1888) = 14.243, $p$ = .001), small effect size ($\varphi$ = .087, $p$ = .001) |
| 1.3 | ($x^2$(2, $N$ = 1591) = 34.355, $p$ < .001), small effect size ($\varphi$ = .147, $p$ < .001) | ($x^2$(1, $N$ = 1590) = 20.396, $p$ < .001), small effect size ($\varphi$ = 113, $p$ < .001) | ($x^2$(2, $N$ = 1554) = 5.948, $p$ = .051), |
| 1.4 | ($x^2$(2, $N$ = 855) = 2.375, $p$ = .305) | ($x^2$(1, $N$ = 855) = 31.287, $p$ < .001), small effect size ($\varphi$ = 191, $p$ < .001) | ($x^2$(2, $N$ = 886) = 1.943, $p$ = .378), |

**Public awareness of search engine optimization (SEO)**

Among internet users, 43% assume a better ranking can be achieved without paying money to Google, as shown in Table 7. Consequently, more than half of the respondents did not think this was possible or said that they did not know (Q2.1). This is in stark contrast to the possibility of influence through paid advertisements, which 79% of internet users are aware of. In two follow-up questions, users who assume the existence of ranking improvements without paying money to Google were asked about both the term "search engine optimization" and SEO tactics. The term to describe these activities ("search engine optimization" or "SEO") is known to 9% of all internet users (Q2.2). In comparison, 15% of users can correctly name SEO techniques (Q2.3) if we consider both correct and partly correct answers as correct.

*Table 7: Public awareness of SEO*

| | Internet users | | Age | | | | | | Educational level | | | | SEO affinity | | | | | |
| | | | under 30 years of age | | 30 to under 50 years of age | | 50 years and older | | A-levels or University degree | | lower | | high | | average | | low | |
| | n | % | n | % | n | % | n | % | n | % | n | % | n | % | n | % | n | % |
|---|---|---|---|---|---|---|---|---|---|---|---|---|---|---|---|---|---|---|
| **Q2.1** | | | | | | | | | | | | | | | | | | |
| Yes | 873 | 43 | 240 | 47 | 387 | 46 | 246 | 38 | 548 | 50 | 326 | 35 | 115 | 61 | 213 | 57 | 589 | 44 |
| No/I don't know | 1138 | 57 | 270 | 53 | 460 | 54 | 408 | 62 | 538 | 50 | 600 | 65 | 73 | 39 | 162 | 43 | 735 | 56 |
| **Q2.2** | | | | | | | | | | | | | | | | | | |
| (Partly) correct | 180 | 9 | 46 | 9 | 97 | 11 | 36 | 6 | 144 | 13 | 36 | 4 | 44 | 23 | 64 | 17 | 117 | 9 |
| Incorrect/I don't know | 692 | 34 | 195 | 38 | 288 | 34 | 210 | 32 | 402 | 37 | 291 | 31 | 72 | 38 | 149 | 40 | 472 | 36 |
| Question not received | 1138 | 57 | 270 | 53 | 460 | 54 | 408 | 62 | 538 | 50 | 600 | 65 | 73 | 39 | 162 | 43 | 735 | 56 |
| **Q2.3** | | | | | | | | | | | | | | | | | | |
| (Partly) correct | 292 | 15 | 84 | 16 | 143 | 17 | 65 | 10 | 223 | 21 | 69 | 7 | 67 | 36 | 104 | 28 | 184 | 14 |
| Incorrect/I don't know | 580 | 29 | 153 | 30 | 244 | 29 | 182 | 28 | 322 | 30 | 256 | 28 | 47 | 25 | 110 | 29 | 404 | 31 |
| Question not received | 1138 | 57 | 270 | 53 | 460 | 54 | 408 | 62 | 538 | 50 | 600 | 65 | 73 | 39 | 162 | 43 | 735 | 56 |

As with the PSM questions, differences between the groups can also be identified for SEO questions. Again, the younger, more highly educated participants and those with an affinity for SEO provided correct answers the most frequently. Correlations between the variables "SEO knowledge" (ratio of correct to incorrect answers for each question) and "groups" (e.g., age groups) are significant for all questions on SEO, with small effect sizes (see Table 8).

*Table 8: Results of the chi-square tests for questions on SEO*

| Question No. | Age (3 groups) | Educational background (2 groups) | SEO affinity (3 groups) |
|---|---|---|---|
| 2.1 | ($\chi^2(2, N = 2011) = 13.499, p = .001$), small effect size ($\varphi = .082, p = .001$) | ($\chi^2(1, N = 2012) = 47.343, p < .001$), small effect size ($\varphi = .153, p < .001$) | ($\chi^2(2, N = 1887) = 30.954, p < .001$), small effect size ($\varphi = .128, p < .001$) |
| 2.2 | ($\chi^2(2, N = 872) = 10.685, p = .005$), small effect size ($\varphi = .111, p = .005$) | ($\chi^2(1, N = 873) = 29.497, p < .001$), small effect size ($\varphi = .184, p < .001$) | ($\chi^2(2, N = 918) = 21.693, p < .001$), small effect size ($\varphi = .154, p < .001$) |
| 2.3 | ($\chi^2(2, N = 871) = 8.190, p = .017$), small effect size ($\varphi = .097, p = .017$) | ($\chi^2(1, N = 870) = 35.386, p < .001$), small effect size ($\varphi = .202, p < .001$) | ($\chi^2(2, N = 916) = 41.775, p < .001$), small effect size ($\varphi = .214, p < .001$) |

Figure 3 contrasts the responses to the questions asked in the two areas of PSM and SEO to illustrate the different levels of knowledge. The percentage of correct answers regarding PSM was considerable higher than it was for SEO.

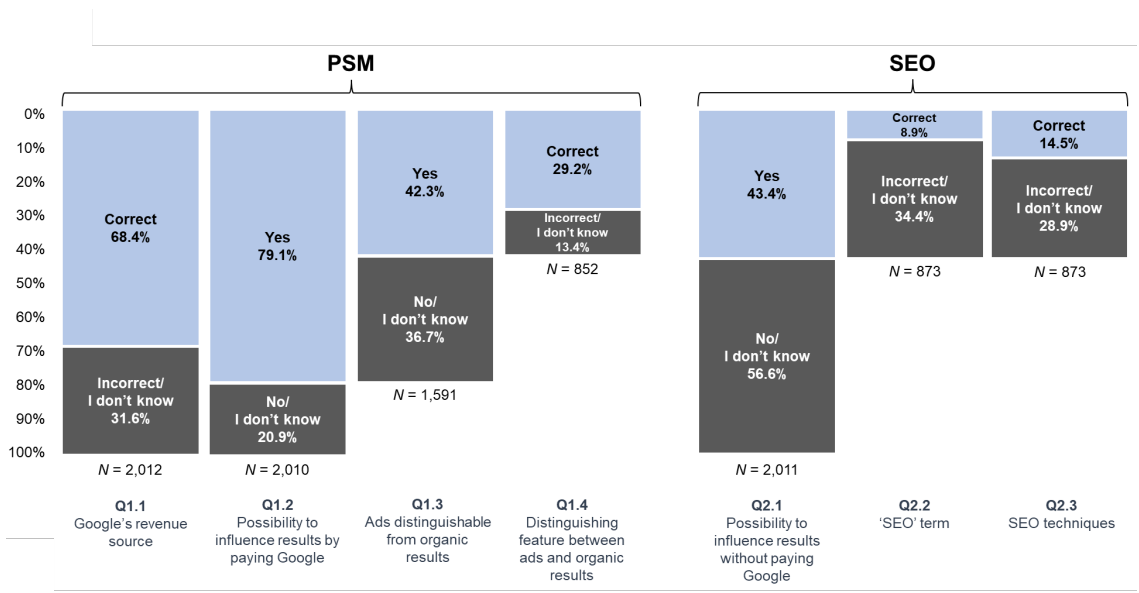

*Figure 3: Public awareness of paid search marketing (PSM) and search engine optimization (SEO)*

## Identification of the areas of influence of SEO and PSM

As we demonstrated in previous sections, internet users have less knowledge of SEO than PSM. The subsequent marking tasks examine the extent to which users can identify search results that are subject to external influence, either via SEO or PSM. The corresponding questions are Q4.1-Q4.4. The success rates were calculated as described in the methods section.

In Figure 4, each bar depicts the average success rate. For example, on the simple structured SERPs (Tasks A/B) on the large screen, subjects achieved a mean success rate of 0.64 for identifying PSM results. The discrepancies with the sample sizes described earlier (large: $N$ = 999, small: $N$ = 1,013) result from erroneous values in the data set, which we excluded before calculating the scores.

Overall, success rates for PSM were significantly higher than for SEO. When comparing the results by SERP complexity level, we can observe that subjects achieved higher success rates on the SERPs with simple structures (tasks A and B) than on the more complex SERPs (tasks C and D). Regarding screen size, SEO results were identified better on the small screen than on the large screen for all tasks. It should be noted here that due to different compositions of search results (see Appendix 1 and Appendix 2), both screen sizes can be compared only to a limited extent. In addition, we can see negative success rates for SEO results on the complex SERPs on the large screen[7]. Hence, more false than correct markings were made in these cases. In summary, it can be said that the possibility of an optimized placement through *SEO* (without paying Google) is considerably less often associated with *organic results* than an optimized placement through *PSM* with *ads*.

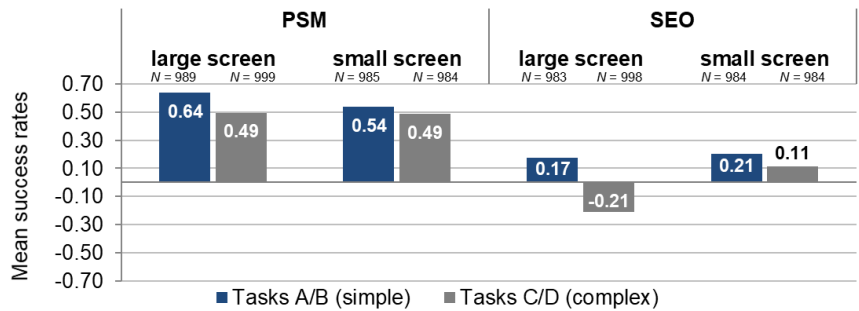

*Figure 4: Mean success rates for marking tasks*

## Public attitudes towards search engine optimization

In addition to questions and tasks regarding knowledge of SEO and PSM, we surveyed the viewpoints of internet users on SEO (Q3.1-3.5, see Study design). For a complete overview of the data for the questions addressed below, see Appendix 3.

*Effect estimations*

In question Q3.1, we asked participants how strong they consider the influence of search engine optimization to be on the ranking of search results. Most assumed that search engine optimization has a (very) strong impact on rankings (59.2% overall). Only a few respondents assumed little (3.1%) or no influence (0.7%). Here, we should keep in mind that an explanation of SEO was provided in advance so that subjects could understand the questions (see methods section).

The responses were quite similar across the user groups. One exception is educational background: more-educated people assume strong or very strong SEO influences significantly more often than less-educated people ($\chi^2(1, N = 2{,}012) = 10.531$, $p = .001$), $\varphi = .072$, $p = .001$.

*Opinions*

In questions Q3.2-3.5, we asked for opinions on SEO. First, we asked participants how large they consider the positive (Q3.2) and negative effects (Q3.3) of SEO to be. Subjects who assumed at least large effects for one of the two questions were then asked in open follow-up questions *which* positive (Q3.4) or negative (Q3.5) effects they assume.

---

[7] Note the method for calculating success rates, which includes the false positives (in the case of SEO results the markings of ads). On the large screen of the complex SERPs, eight shopping ads were displayed, and only two on the small screen. Thus, it was more difficult on the large screen to accurately identify SEO results because more ads were shown. The differences between the screen sizes in the marking of the SEO results on the complex SERPs are therefore probably also due to the different number of ads.

Results show that search engine optimization is more often perceived as positive (75.2%) than negative (68.4%) when responses for "medium," "large," and "very large" are considered together, as shown in Figure 5. We found only one significant difference between the groups. Those with higher levels of education assume negative SEO effects significantly more often than less-educated individuals ($\chi^2$(1, *N* = 2,011) = 9.253, *p* = .002), *φ* = .068, *p* = .002.

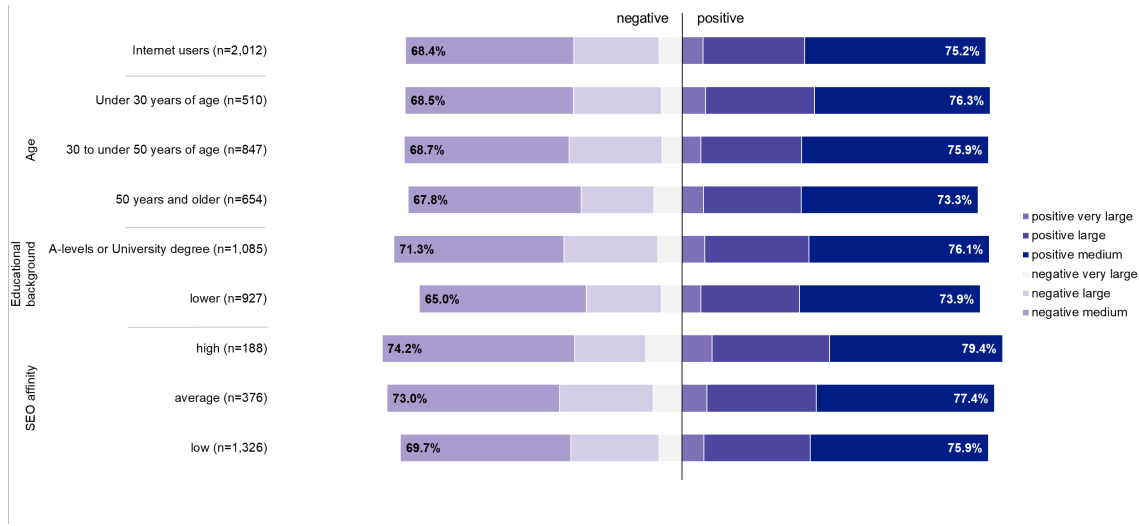

*Figure 5: Opinions on SEO; group comparison for Q3.2 and Q3.3. Numbers indicate users who assume at least a medium effect*

Looking at the assumed positive and negative SEO influences together (Table 9), we can see that a large group of respondents (27.4%) assumes both medium positive and medium negative SEO influences at the same time. About every tenth internet user (9.6%) responds that they do not know the influence in either case. Furthermore, more subjects (1.7%) indicate both very large positive *and* very large negative SEO influences than indicate very large positive *without* any negative SEO influences (0.5%).

*Table 9: Cross table of positive and negative SEO opinions*

|  |  | positive SEO influence (%) | | | | | | % of users |
|---|---|---|---|---|---|---|---|---|
|  |  | I don't know | non-existent | low | medium | large | very large |  |
| **negative SEO influence (%)** | I don't know | 9.6 | 0.0 | 0.3 | 0.7 | 0.5 | 0.2 | 0.0 |
|  | non-existent | 0.0 | 0.5 | 0.2 | 0.5 | 0.4 | 0.5 | 0.1-0.5 |
|  | low | 0.0 | 0.1 | 1.9 | 7.3 | 7.7 | 1.1 | 0.6-1.0 |
|  | medium | 0.4 | 0.3 | 3.1 | 27.4 | 9.7 | 0.8 | 1.1-5.0 |
|  | large | 0.3 | 0.6 | 4.6 | 8.4 | 6.4 | 0.6 | 5.1-10.0 |
|  | very large | 0.1 | 1.3 | 1.2 | 0.7 | 0.5 | 1.7 | 10.1- |

We coded the open answers to the sub-question on user opinion about SEO. By the time we asked participants for their opinions, they all had read the short text explaining SEO. Among n=297 positive SEO opinions, the most common response (Q3.4) was better result quality (40%), followed by quicker retrieval (25%), advantages for website operators (18%), and advantages for the searcher such as individualization and filters (15%). Among n=320 negative responses, the most frequently mentioned opinion about SEO (Q3.5) was manipulating results with negative intentions (32%), followed by a displacement of content actually being searched for (30%), negative influences on results quality (19%), and discrimination against smaller websites (18%)[8].

## Discussion

Our study reveals a low level of knowledge regarding external influences on SERPs, which we found both in the answers to the knowledge-based and the task-based questions. The lack of knowledge was most profound with respect to the field of search engine optimization (SEO). The possibility of influence through paid results (PSM) is known to significantly more internet users than SEO, i.e., the possibility of improving rankings without paying Google (79% vs. 43%). The results of the marking tasks are in line with these findings. Ads were identified comparatively well (every second ad on average), but only one out of ten organic results were associated with SEO. Thus, we find that users are largely unable to distinguish between results that can be influenced through search engine optimization and results that are directly paid for. Most users assume a strong influence of SEO on search results and see both positive and negative effects of this influence. Thus, from the user's point of view, the advantage of obtaining better quality results is contrasted with the disadvantage that influence with negative intentions could occur through SEO. Furthermore, we identified significant differences in age, level of education, and affinity for SEO topics. Younger, well-educated people and those whose professions or studies suggested a familiarity with SEO topics were most likely to provide correct answers on the questionnaire.

Compared to the results from Lewandowski et al. (2018), a similar number of users know that ads differ from organic results (42% in the present study vs. 43% in the older study). Still, fewer internet users know *how* ads differ from organic results (29% vs. 34%) and where Google's revenue comes from (68% vs. 81%). However, more people indicated they know there is a paid placement option on SERPs (79% vs. 73%).

It is paramount for users to understand the information retrieval systems they are using, even more so when commercial interests can influence results. Our study shows that users are not well-informed about these influences. Yet this knowledge is essential for being a digitally empowered citizen. It is clear that media and information literacy interventions can have a positive effect on the ability to critically examine online content and judge the credibility of online news (Guess et al., 2020). We advocate for information literacy education that teaches people the best ways to use search engines and how external actors can influence the results these engines provide. However, one should keep in mind that such interventions place responsibility in the hands of users while ignoring the responsibility of the search engine providers (Lewandowski, 2017a). The processes that lead users to select a particular result are complex and lie not only with the user but also with the search engine provider and external parties aiming to influence the results.

Of course, this study has its limitations. Due to time restrictions, we were only able to use two marking tasks per participant. Nevertheless, we were able to increase the number of SERPs investigated by randomly dividing the sample into two groups, leading to four SERPs examined. Still, it would be desirable to use more SERPs with a larger number of result types (e.g., knowledge-graph results and different vertical results) to get a more differentiated picture of user abilities with respect to identifying PSM and SEO respectively. We targeted the research questions using an online questionnaire containing marking tasks. As a result, our study is not based on real interactions between users and search engines. More interactive studies, e.g., in lab settings, could enrich and differentiate the results of our online survey.

---

[8] We could not assign 2% of the positive and 1% of the negative responses to a category since the statements were unclear.

While a sample representative of the German online population is a solid basis for strong empirical evidence, it is geographically limited. It would be interesting to repeat the study in other countries to strengthen the empirical evidence of the results. Although we expect very similar findings internationally, results could differ somewhat from one country to the next. Furthermore, repeating the survey every year would allow for long-term monitoring of search engine use, something that is currently not being done. A more interactive environment would make it possible to analyze search sessions, where users search for information and are confronted with PSM and SEO results as they search. Whereas it is easy to model PSM results, this does not hold for SEO results. Research has found that a significant portion of results displayed by Google are optimized (Lewandowski et al., 2021). Consequently, one would need to manipulate the results shown so users also see non-optimized or only lightly optimized results. Otherwise, users might come across optimized results only. A very fruitful way to investigate user interactions with search engines would be to use real usage data from commercial search engine providers. Such data is collected by these companies but is generally not available to researchers. This problem has often been lamented (e.g., Lazer et al., 2020), and it has been argued, for instance, that researchers should build a "Human Screenome" by using logging software extensively to get an understanding of how people use digital media (Reeves et al., 2020).

Our study has shown that the search engine optimization industry operates largely outside the awareness of search engine users, even though it potentially has an enormous influence on the results users see and which sites they choose to visit. This brings us back to the points raised in the introduction: Search engine optimization not only influences which products and services appear at the top of search engine result pages, but also what content users see when they research topics of societal relevance. Like any other actor using SEO methods, actors distributing misinformation will profit immensely if they manage to make their content visible in the top results of search engines, as they will not only gain visibility but also profit from the trust users place in search engines, Google in particular. However, the effect that these actors have on search results has not yet been quantified.

An important question is how providers of credible information should deal with competition from sites that disseminate misinformation or inaccurate information. A simple solution would be for legitimate sites to use search engine optimization measures to promote their content. They may even have an advantage here because search engines consider source credibility in their ranking algorithms. We fear, however, that the public sector, non-governmental organizations, and other non-profits will not be able to compete with big companies pursuing commercial interests. Furthermore, this would lead to an arms race between these different actors. In the face of such limited options, the question of regulation arises. Several proposals for regulating information gatekeepers such as search engines have been made in the past (e.g., Grimmelmann, 2009). When such approaches deal with search results at all, they focus on targeted manipulation of results by the search engine companies (Vogl & Barrett, 2010), not external manipulations or user understanding of search results.

Search engine companies are in fact already working towards making search results more diverse (e.g., Mitchell et al., 2020) by showing a selection of results from different sources or arguing for a different viewpoint in the top results. While we welcome such efforts, they can target the problem described in this paper only indirectly: While some results might be demoted, SEO will still influence which results are shown for particular aspects of a topic. This leads to the open question of how external influences on search results such as SEO can be addressed so that they lead to a fair and transparent competition for visibility. Furthermore, increasing the diversity of organic results will not affect the influence of advertisements since most users do not know that ads differ from organic results (58%) and how they differ (71%), as described in this paper.

For future research, we suggest that the quantitative analysis presented here should be complemented by qualitative studies that focus on gaining a deeper understanding of user knowledge regarding influences on SERPs and attitudes about how commercial search engines present results and lead users to select certain ones. A very important question not addressed in the current study is how search engine optimization influences result relevance. While a positive effect of SEO can be that it brings relevant results to the top of result lists, it could also have a negative effect by increasing the rankings of less-relevant results at the expense of more relevant but less-optimized content. Another question that should be addressed in future research is how the interests of search engine providers influence which results users select and thus the information they consider when making decisions. Further, with regard to the multiple external influences on search results, future research should measure these influences more extensively, including their interplay. For instance, are the influences cumulative in their impact, or do they possibly even multiply the effect of external influence? For instance, when content from one source takes up more "screen real estate" (Nicholson et al., 2006) (an ad plus some organic and vertical results on the same result page, for example), does this lead users to assign more credibility to that source?

## Conclusion

In this paper, we investigated the role of search engine optimization (SEO) from a user perspective by conducting a large-scale, representative online survey with $N$ = 2,012 German internet users. We examined search engine optimization within the context of external influences on search results in commercial search engines. One reason why search engine optimization is so influential and can lead to tremendous growth in traffic is that there are only a few relevant search engines, with Google dominating most markets. We found that users do not know why specific results are shown on SERPs. Many cannot differentiate between results that can be influenced through paid search marketing (PSM) and search engine optimization (SEO).

This paper contributes to information science research on search engines as we show empirically how users dependent on commercial search engines are left with results generated from a mixture of relevance criteria, search engine provider self-interests, and external influences. This leads us to call for a broad discussion on how search systems should be designed to best serve the needs of society. From our results, it seems clear that leaving knowledge acquisition on the web in the hands of only one or a few commercial search engines is a problematic scenario. It is even more so when we consider misinformation. One proposed solution is to increase the number of search engines by supporting search engine development with public funding of an Open Web Index (Lewandowski, 2019) upon which new search engines can be built.

## Acknowledgments

The online survey was conducted by Fittkau & Maaß Consulting. Research data is available at OSF; https://doi.org/10.17605/OSF.IO/JYV9R. A data note of the online survey is available at F1000 Research; https://doi.org/10.12688/f1000research.109662.1.

## Funding

This work was funded by the German Research Foundation (DFG – Deutsche Forschungsgemeinschaft; Grant No. 417552432).

# Appendix 1

SERP screenshots for tasks A-D, large screen. PSM results: dashed line, SEO results: solid line.

**Task A**

**Task B**

**Task C**

**Task D**

# Appendix 2

SERP screenshots for tasks A-D, small screen. PSM results: dashed line, SEO results: solid line.

| Task A | Task B | Task C | Task D |
|---|---|---|---|
| 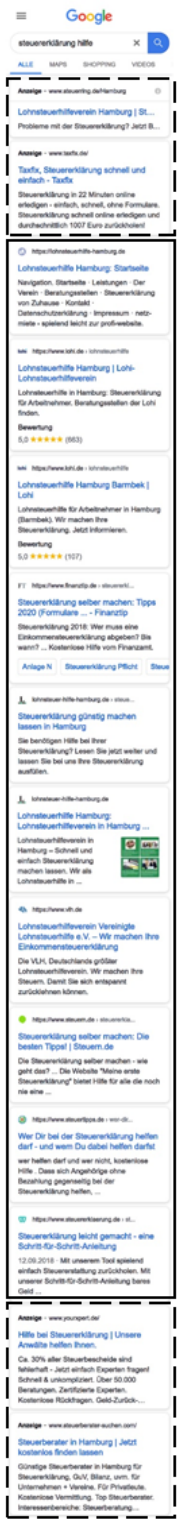 | 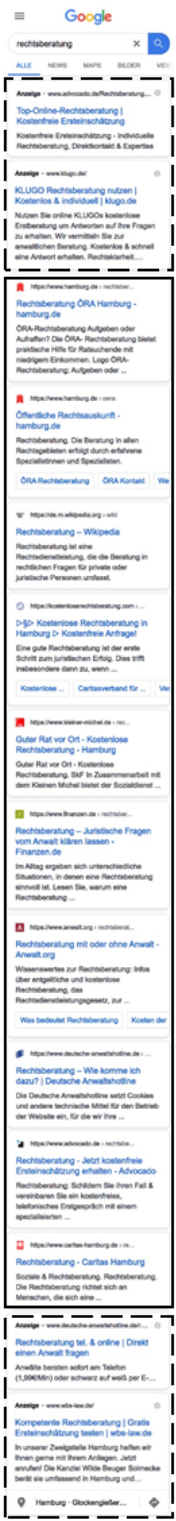 | 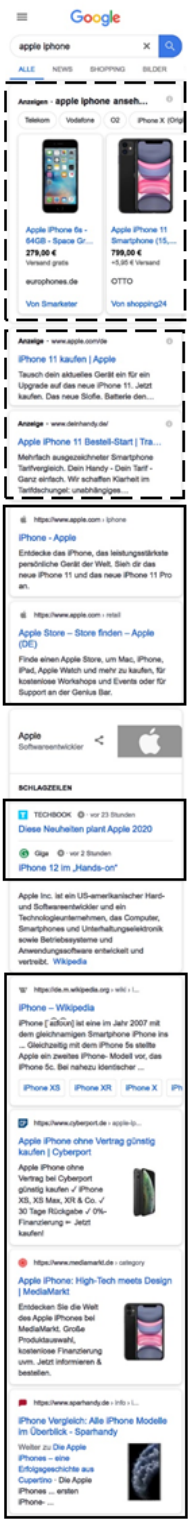 | 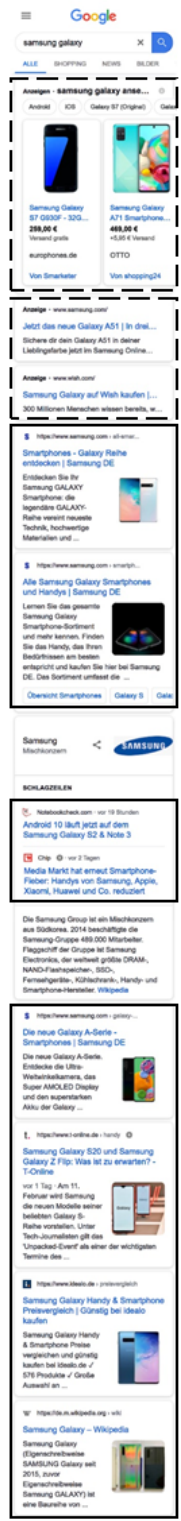 |

# Appendix 3

*Table A1: Public attitudes towards search engine optimization*

| | Internet users | | Age | | | | | | Educational level | | | | SEO affinity | | | | | |
| | | | under 30 years of age | | 30 to under 50 years of age | | 50 years and older | | A-levels or University degree | | lower | | high | | average | | low | |
| | n | % | n | % | n | % | n | % | n | % | n | % | n | % | n | % | n | % |
| **Q3.1** | | | | | | | | | | | | | | | | | | |
| very strong influence | 348 | 17 | 77 | 15 | 145 | 17 | 126 | 19 | 197 | 18 | 151 | 16 | 36 | 19 | 74 | 20 | 236 | 18 |
| major influence | 842 | 42 | 219 | 43 | 353 | 42 | 271 | 41 | 481 | 44 | 361 | 39 | 85 | 45 | 164 | 44 | 561 | 42 |
| medium influence | 585 | 29 | 170 | 33 | 240 | 28 | 175 | 27 | 317 | 29 | 268 | 29 | 55 | 29 | 110 | 29 | 387 | 29 |
| little influence | 63 | 3 | 15 | 3 | 25 | 3 | 23 | 4 | 29 | 3 | 35 | 4 | 6 | 3 | 8 | 2 | 38 | 3 |
| no influence | 15 | 1 | 3 | 1 | 8 | 1 | 4 | 1 | 7 | 1 | 8 | 1 | 2 | 1 | 4 | 1 | 9 | 1 |
| I don't know | 158 | 8 | 25 | 5 | 77 | 9 | 56 | 9 | 54 | 5 | 104 | 11 | 5 | 3 | 16 | 4 | 93 | 7 |
| **Q3.2** | | | | | | | | | | | | | | | | | | |
| very large | 104 | 5 | 30 | 6 | 40 | 5 | 35 | 5 | 61 | 6 | 43 | 5 | 14 | 8 | 23 | 6 | 72 | 5 |
| large | 506 | 25 | 137 | 27 | 211 | 25 | 159 | 24 | 280 | 26 | 226 | 24 | 55 | 29 | 101 | 27 | 350 | 26 |
| medium | 900 | 45 | 222 | 44 | 392 | 46 | 286 | 44 | 485 | 45 | 415 | 45 | 80 | 43 | 166 | 44 | 584 | 44 |
| low | 233 | 12 | 67 | 13 | 92 | 11 | 73 | 11 | 143 | 13 | 90 | 10 | 27 | 14 | 50 | 13 | 155 | 12 |
| non-existent | 57 | 3 | 12 | 2 | 20 | 2 | 25 | 4 | 36 | 3 | 21 | 2 | 5 | 3 | 11 | 3 | 38 | 3 |
| I don't know | 210 | 10 | 41 | 8 | 92 | 11 | 76 | 12 | 80 | 7 | 130 | 14 | 7 | 4 | 25 | 7 | 126 | 10 |
| **Q3.3** | | | | | | | | | | | | | | | | | | |
| very large | 113 | 6 | 26 | 5 | 41 | 5 | 45 | 7 | 65 | 6 | 47 | 5 | 17 | 9 | 27 | 7 | 74 | 6 |
| large | 423 | 21 | 110 | 22 | 195 | 23 | 118 | 18 | 251 | 23 | 172 | 19 | 33 | 18 | 87 | 23 | 290 | 22 |
| medium | 839 | 42 | 212 | 42 | 346 | 41 | 281 | 43 | 456 | 42 | 382 | 41 | 89 | 48 | 160 | 43 | 559 | 42 |
| low | 367 | 18 | 105 | 21 | 149 | 18 | 114 | 17 | 205 | 19 | 162 | 18 | 36 | 19 | 67 | 18 | 240 | 18 |
| non-existent | 44 | 2 | 10 | 2 | 13 | 2 | 21 | 3 | 21 | 2 | 24 | 3 | 6 | 3 | 8 | 2 | 26 | 2 |
| I don't know | 225 | 11 | 45 | 9 | 104 | 12 | 76 | 12 | 86 | 8 | 139 | 15 | 7 | 4 | 27 | 7 | 136 | 10 |
| **Q3.4** | | | | | | | | | | | | | | | | | | |
| free input | 297 | 49 | 90 | 54 | 127 | 51 | 80 | 42 | 184 | 54 | 113 | 42 | 48 | 71 | 70 | 56 | 204 | 48 |
| I don't know | 314 | 51 | 77 | 46 | 124 | 49 | 113 | 58 | 158 | 46 | 156 | 58 | 20 | 29 | 55 | 44 | 217 | 52 |
| **Q3.5** | | | | | | | | | | | | | | | | | | |
| free input | 320 | 60 | 80 | 59 | 145 | 61 | 97 | 60 | 204 | 64 | 118 | 54 | 35 | 70 | 74 | 65 | 208 | 57 |
| I don't know | 216 | 40 | 57 | 41 | 92 | 39 | 66 | 40 | 113 | 36 | 101 | 46 | 15 | 30 | 40 | 35 | 156 | 43 |